\documentclass{elsart}
\usepackage{graphicx}

\def\textbf#1{{\bf #1}}
\def\textit#1{{\it #1}}

\begin{document}

\begin{frontmatter}
  
\title{Dynamics of price and trading volume in a spin model of stock
  markets with heterogeneous agents}
\author[Tokyo]{Taisei Kaizoji\thanksref{contact}},
  \ead{kaizoji@icu.ac.jp}
\author[Kiel]{Stefan Bornholdt}
\author[Nara]{Yoshi Fujiwara}
\address[Tokyo]{%
  Division of Social Sciences,
  International Christian University, Osawa, Mitaka, Tokyo 181-8585,
  Japan}
\address[Kiel]{%
  Institute for Theoretical Physics,
  University of Kiel, Leibnizstrasse 15, D-24098 Kiel, Germany}
\address[Nara]{%
  Keihanna Research Center,
  Communications Research Laboratory, \\Kyoto 619-0289, Japan}
\thanks[contact]{%
  Corresponding author. Division of Social Sciences,
  International Christian University, Osawa, Mitaka, Tokyo 181-8585,
  Japan. FAX: +81-422-33-3367.
}

\begin{abstract}
  The dynamics of a stock market with heterogeneous agents is
  discussed in the framework of a recently proposed spin model for the
  emergence of bubbles and crashes. We relate the log returns of stock
  prices to magnetization in the model and find that it is closely
  related to trading volume as observed in real markets. The
  cumulative distribution of log returns exhibits scaling with
  exponents steeper than $2$ and scaling is observed in the
  distribution of transition times between bull and bear markets.
\end{abstract}


\end{frontmatter}
\section{Introduction}

During the last years there has been great interest in applications of
statistical physics to financial market dynamics. A variety of
agent-based models have been proposed over the last few years to study
financial market dynamics
\cite{Bak97,Challet97,Cont00,Lux99,Stauffer99}. In particular, spin
models as the most popular models of statistical mechanics have been
applied to describe the dynamics of traders in financial markets by
several researchers
\cite{Bornholdt01,Chowdhury99,Iori99,Kaizoji00,Roehner00,Weron00,Yamano02}.

A particularly simple model of a stock market in the form of a spin
model motivated by the Ising model has been proposed recently
\cite{Bornholdt01}, in order to simulate the dynamics of expectations
in systems of many agents. The model introduces a new coupling that
relates each spin (agent) to the global magnetization of the spin
model, in addition to the ferromagnetic (Ising) couplings connecting
each spin to its local neighborhood. The global coupling effectively
destabilizes local spin orientation, depending on the size of
magnetization. The resulting frustration between seeking ferromagnetic
order locally, but escaping ferromagnetic order globally, causes a
metastable dynamics with intermittency and phases of chaotic dynamics.
In particular, this occurs at temperatures below the critical
temperature of the Ising model. While the model exhibits dynamical
properties which are similar to the stylized facts observed in
financial markets, a careful interpretation in terms of financial
markets is still lacking. This is the main goal of this paper. In
particular, \cite{Bornholdt01,Yamano02} treat the magnetization of the
model as price signal which, however, is unnatural when deriving a
logarithmic return of this quantity as these authors do. As a result,
small magnetization values cause large signals in the returns with an
exponent of the size distribution different from the underlying
model's exponent.

Let us here consider Bornholdt's spin model in the context of a stock
market with heterogeneous traders. The aim of this paper is (i) to
interpret the magnetization of the spin model in terms of financial
markets and to study the mechanisms that create bubbles and crashes,
and (ii) to investigate the statistical properties of market price and
trading volume. The new elements in the model studied in this paper
are the explicit introduction of two groups of traders with different
investment strategies, fundamentalists and interacting
traders\footnote{The interacting traders are often called noise
  traders in finance literature. }, as well as of a market clearing
system that executes trading at matched book. Given these conditions,
the market price is related to the sum of fundamental price and
magnetization, and the trading volume is simply given by the
magnetization. We also show that the model is able to explain the
empirically observed positive cross-correlation between volatility and
trading volume\footnote{The positive cross-correlation between
  volatility and trading volume is demonstrated by \cite{Gallant92}.
  However, little attention has been paid to the relationship between
  price and trading volume in terms of theory, Iori \cite{Iori01}
  being among those who first studied this relationship.}. Finally, we
observe that the model reproduces the well-known stylized facts of the
return distribution such as fat tails and volatility clustering
\cite{Gopi00a,Liu99}, and study volatilities at different time-scales.

\section{The Model}

Let us consider a stock market where a large stock is traded at price
$p(t)$. Two groups of traders with different trading strategies,
\textit{fundamentalists} and \textit{interacting traders}, participate
in the trading. The number of fundamentalists $m$ and the number of
interacting traders $n$ are assumed to be constant. The model is
designed to describe the stock price movements over short periods,
such as one day. In the following, a more precise account of the
decision making of each trader type is given.

\subsection{Fundamentalists}

Fundamentalists are assumed to have a reasonable knowledge of the
fundamental value of the stock $p^*(t)$. If the price $p(t)$ is below
the fundamental value $ p^*(t) $, a fundamentalist tends to buy the
stock (as he estimates the stock to be undervalued), and if the price
is above the fundamental value, a fundamentalist tends to sell the
stock (then considered as an overvalued and risky asset). Hence we
assume that fundamentalists' buying or selling order is given by:
\begin{equation}
x^F(t) = a\; m\; \left(\ln p^*(t) - \ln p(t)\right),
\label{eqn:a1}
\end{equation}
where $m$ is the number of fundamentalists, and $a$ parametrizes the
strength of the reaction on the discrepancy between the fundamental
price and the market price.

\subsection{Interacting Traders}

During each time period, an interacting trader may choose to either
buy or sell the stock, and is assumed to trade a fixed amount of the
stock $b$ in a period of trading. Interacting traders are labeled by
an integer $1 \leq i \leq n$. The investment attitude of interacting
trader $i$ is represented by the random variable $s_i$ and is defined
as follows: If interacting trader $i$ is a buyer of the stock during a
given period, then $s_{i} = + 1$, otherwise he sells the stock and
$s_{i} = - 1$.

Now let us formulate the dynamics of the investment attitude of
interacting traders in terms of the spin model \cite{Bornholdt01}. Let
us consider that the investment attitude $s_{i}$ of interacting trader
$i$ is updated with a heat-bath dynamics according to
\begin{eqnarray}
 s_i(t+1) & = & + 1 \quad\mbox{with}\quad p = \frac{1}{1 + \exp(-2 \beta h_i(t))} \\ \label{eqn:a2}
 s_i(t+1) & = & - 1 \quad\mbox{with}\quad 1 - p 
\label{eqn:a3}
\end{eqnarray}
where $ h_i(t) $ is the local field of the spin model, governing the
strategic choice of the trader.

Let us consider the simplest possible scenario for local strategy
changes of an interacting trader. We assume that the decision which an
interacting trader makes is influenced by two factors, local
information, as well as global information. Local information is
provided by the nearby interacting traders' behavior. To be definite,
let us assume that each interacting trader may only be influenced by
its nearest neighbors in a suitably defined neighborhood. Global
information, on the other hand, is provided by the information whether
the trader belongs to the majority group or to the minority group of
sellers or buyers at a given time period, and how large these groups
are.  The asymmetry in size of majority versus minority groups can be
measured by the absolute value of the magnetization $ |M(t)| $, where
\begin{equation}
 M(t) = \frac{1}{n} \sum^n_{i=1} s_i(t).    
\label{eqn:a4}
\end{equation} 

The goal of the interacting traders is to obtain capital gain through
trading. They know that it is necessary to be in the majority group in
order to gain capital, however, this is not sufficient as, in
addition, the majority group has to expand over the next trading
period. On the other hand, an interacting trader in the majority group
would expect that the larger the value of $ |M(t)| $ is, the more
difficult a further increase in size of the majority group would be.
Therefore, interacting traders in the majority group tend to switch to
the minority group in order to avert capital loss, e.g.\ to escape a
large crash, as the size of the majority group increases. In other
words, the interacting trader who is in the majority group tends to be
a risk averter as the majority group increases. On the other hand, an
interacting agent who is in the minority group tends to switch to the
majority group in order to gain capital. An interacting agent in the
minority group tends to be a risk taker as the majority group
increases.

To sum up, the larger $ |M(t)| $ is, the larger the probability with
which interacting traders in the majority group (interacting traders
in the minority group, respectively) withdraw from their coalition.
Following \cite{Bornholdt01}, the local field $ h_i(t) $ containing
the interactions discussed above is specified by
\begin{equation}
 h_i(t) = \sum^m_{j=1} J_{ij} S_j(t) - \alpha S_j(t) |M(t)| 
\label{eqn:a5}
\end{equation}
with a global coupling constant $ \alpha > 0 $. The first term is
chosen as a local Ising Hamiltonian with nearest neighbor interactions
$ J_{ij} = J $ and $ J_{ii} = 0 $ for all other pairs.

We assume that the interacting-traders' excess demand for the stock is
approximated as
\begin{equation} 
   x^I(t) = b\; n\; M(t). 
\label{eqn:a6}
\end{equation}

\subsection{Market price and trading volume}

Let us leave the traders' decision-making processes and turn to the
determination of the market price. We assume the existence of a market
clearing system. In the system a \textit{market maker} mediates the
trading and adjusts the market price to the market clearing values.
The market transaction is executed when the buying orders are equal to
the selling orders.

The balance of demand and supply is written as
\begin{equation}
 x^F(t)+ x^I(t) = a\; m\; \left[\ln p^*(t) - \ln p(t)\right] + b\; n\; M(t) = 0.  
\label{eqn:a7}
\end{equation}
Hence the market price and the trading volume are calculated as 
\begin{equation}
\ln p(t) = \ln p^*(t) + \lambda M(t), \quad \lambda = \frac{b\; n}{a\; m}, 
\label{eqn:a8}
\end{equation}
and 
\begin{equation}
V(t) = b\; n\; \frac{1+|M(t)|}{2}. 
\label{eqn:a9}
\end{equation} 
Using the price equation (8) we can categorize the market situations
as follows: If $ M(t) = 0 $, the market price $ p(t) $ is equal to the
fundamental price $ p^*(t) $. If $ M(t) > 0 $, the market price $ p(t)
$ exceeds the fundamental price $ p^*(t) $ ({\it bull} market regime).
If $ M(t) < 0 $, the market price $ p(t) $ is less than the
fundamental price $ p^*(t) $ ({\it bear} market regime). Using (8),
the logarithmic relative change of price, the so-called log-return, is
defined as
\begin{equation}
 \ln p(t) - \ln p(t-1) = (\ln p^*(t) - \ln p^*(t-1)) + \lambda (M(t) - M(t-1)). 
\label{eqn:a10}
\end{equation}
Let us consider for a moment that only fundamentalists participate in
trading. Then in principle the market price $p(t)$ is always equal to
the fundamental price $ p^*(t) $. This implies that the so-called {\it
  efficient market hypothesis} holds. Following the efficient market
model by \cite{Fama70} then the fundamental price $p^*(t)$ follows a
random walk. Since the continuous limit of a random walk is a Gaussian
process, the probability density of the log-return, defined as $r(t) =
\ln p(t) - \ln p(t-1)$, is normal. For real financial data, however,
there are strong deviations from normality\footnote{Furthermore
  financial asset prices are too volatile to accord with efficient
  markets as has been demonstrated by \cite{LeRoy81} and
  \cite{Shiller81}.}. As we discuss here, including both
fundamentalists and interacting traders to coexist in the market,
offers a possible mechanism for excessive fluctuations such as bull
markets and bear markets. 

To investigate the statistical properties of the price and the trading
volume in the spin model of stock markets, we will assume for
simplicity that the fundamental price is constant over time.

\section{Simulations} 

\subsection{Bubbles and crashes} 

In the new framework developed so far we see that the dynamics of the
log-return corresponds to the linear change in absolute magnetization
of the spin model \cite{Bornholdt01}. Typical behavior of the such
defined return $r(t)$ as well as the magnetization $M(t)$ is shown in
Figures 1(a) and 1(b). Here, a 101*101 lattice of the general version
of the model as defined in eq. (8) and with the spins updated
according to (5) is shown. It is simulated at temperature
$T=1/\beta=0.5$ with couplings $J=1$ and $\alpha=20$, using random
serial and asynchronous heat bath updates of single sites. In Figure
1(a) the intermittent phases of ordered and turbulent dynamics of the
log-return are nicely visible. Qualitatively, this dynamical behavior
is similar to the dynamics of daily changes of real financial indices,
as for example the Dow Jones Index shown in Figure 1(c). To some
degree, these transitions of the return can be related to the
magnetization in the spin model. Figure 1 (b) illustrates that the
bull (bear) market $ M(t) > 0$ ($M(t)<0$) becomes unstable, and the
transition from a metastable phase to a phase of high volatility
occurs, when the absolute magnetization $ |M(t)| $ approaches some
critical value. Noting that trading volume is defined as $b\; n\;
(1+|M(t)|)/2$, this suggests that also some critical trading volume
exists near the onset of turbulent phases. This is in agreement with
the empirical study by \cite{Gallant92} who found the empirical
regularities: (i) positive correlation between the volatility and the
trading volume; (ii) large price movements are followed by high
trading volume.

The origin of the intermittency can be seen in the local field $ h_i $
eq.\ (5) representing the external influences on the decision-making
of interacting-trader $i$. In particular, the second term of $ h_i $
tends to encourage a spin flip when magnetization gets large. Thus
each interacting-trader frequently switches his strategy to the
opposite value if the trading volume gets large. As a consequence, the
bull (bear) market is unstable and the phase of the high volatility
appears. The metastable phases are the analogue of speculative bubbles
as, for example, the bull market is defined as a large deviation of
the market price from the fundamental price. In fact it is a common
saying that ``it takes trading volume to move stock price'' in the
real stock market. A typical example is the crash of Oct.\ 1987, when
the Dow Jones Industrial Average dropped 22.6\% accompanied by an
estimated $ 6 \times 10^8 $ shares that changed hands at the New York
Stock Exchange alone \cite{Gopi00b}. Fama \cite{Fama89} has argued
that the crash of Oct.\ 1987 at the US and other stock markets
worldwide could be seen as the signature of an efficient reassessment
of and convergence to the correct fundamental price after the long
speculative bubble proceeding it.

It is interesting to investigate how long the bull or bear markets
last from the point of view of practical use. Figure 2 shows the
distribution of the bull (bear) market durations that is defined as
the period from the beginning of a bull (bear) market $M(t) \geq 0 \;
(M(t) \leq 0)$ to the end of the bull (bear) market $M(t)=0$. In other
words, the bull (bear) market duration means the period from a point
of time that the market price exceeds the fundamental price to the
next point of time that the market price falls short of the
fundamental price. In the model one observes that the bull (or bear)
market durations are power law distributed with an exponent of
approximately $-1.3$.

\subsection{Fat tails and volatility clustering}

As shown in the previous works \cite{Bornholdt01} and \cite{Yamano02}
the simple spin model considered here reproduces major stylized facts
of real financial data. The actual distribution of log-return $r(t)$
has {\it fat tails} in sharp contrast to a Gaussian distribution
(Figure 3).  That is, there is a higher probability for extreme values
to occur as compared to the case of a Gaussian distribution. Recent
studies of the distribution for the absolute returns $|r(t)|$ report
power law asymptotic behavior,
\begin{equation}
    P(|r(t)| > x) \sim \frac{1}{x^\mu}
\label{eqn:a11}
\end{equation}
with an exponent $\mu$ between about 2 and 4 for stock returns. Figure
4 shows the cumulative probability distribution of the absolute return
that is generated from the model. The observed model exponent of $\mu
= 2.3$ lies in the range of empirical data.

Numerous empirical studies show that the volatility $ |r(t)| $ on
successive days is positively correlated, and these correlations
remain positive for weeks or months. This is called clustered
volatility. Furthermore the autocorrelation function for volatility
decays slowly, and sometimes a power law decay is observed. As seen in
Figure 1(a), phases of high volatility in the model dynamics are
strongly clustered. The corresponding autocorrelation of volatility $
|r(t)| $ is shown in Figure 5, with volatility clustering on a
qualitatively similar scale as observed in real financial market data.

\subsection{Volatilities at different time-scales} 

Let us consider a time-scale $\tau$ at which we observe price
fluctuations. The log-return for duration $\tau$ is then defined as
$r_{\tau}(t)=\ln(P(t)/P(t-\tau))$. Volatility clustering as described
in the previous section is this observable defined for an interval
$\tau$ ranging from several minutes to more than a month or even
longer. In this intermediate and strongly correlated regime of
time-scales, volatilities at different time-scales may show
self-similarity \cite{Mandelbrot97,Muller97,Arnedo98,Fujiwara01}.

Self-similarity in this context states that volatilities
$v_{\tau}\equiv|r_{\tau}|$ at different time-scales $\tau$ are related
in such a way that that the ratio of volatilities at two different
scales does not statistically depend on the coarse-graining level.
Thus daily volatility is related to weekly, monthly volatilities by
stochastic multiplicative factors. The self-similarity has been shown
to be equivalent to scaling of moments under some conditions
\cite{Mandelbrot97}\cite{Muller97}\cite{Arnedo98}.  Scaling occurs
when $\left\langle\ {v_{\tau}{}^q}\ \right\rangle\propto t^{\phi(q)}$,
where $\phi(q)$ is the scaling function which is related to the
statistical property of the multiplicative factors \cite{Fujiwara01}.
Figure 6 (a) depicts the presence of self-similarity in actual data of
the NYSE stock index.

Though it is not straightforward to relate time-scales between the
simulation and real data, it is interesting to look at how the
volatilities at different time-scales behave in the regime where the
volatility clustering is valid. Figure 6 (b) shows the scaling of
moments from the data of log-returns calculated for different scales
in time-steps of the simulation. We observe a range where
self-similarity dominates and that this property is broken at some
time-scale, which corresponds to the scale where volatility clustering
as seen in Figure 5 deviates from a power-law of the autocorrelation
function.  This observation is encouraging as it might help relate the
time-scales of simulations and real markets.

\section{Concluding remarks} 

In this paper we have considered the spin model presented in
\cite{Bornholdt01} in the context of a stock market with heterogeneous
traders, that is, fundamentalists and interacting traders. We have
demonstrated that magnetization in the spin model closely corresponds
to trading volume in the stock market, and the market price is
determined by magnetization under natural assumptions. As a
consequence we have been able to give a reasonable interpretation to
an aperiodic switching between bull markets and bear markets observed
in Bornholdt's spin model.  As a result, the model reproduces main
observations of real financial markets as power-law distributed
returns, clustered volatility, positive cross-correlation between
volatility and trading volume, as well as self-similarity in
volatilities at different time-scales.  We also have found that
scaling is observed in the distribution of transition times between
bull and bear markets. Although the power law scaling of the
distribution has never been examined empirically on short time scales,
the power law statistics showed here is not only an interesting
finding theoretically but presumably also useful to measure the risk
of security investments in practice.  The empirical study will be left
for future work.


\newpage

\begin{figure}
\includegraphics[height=21cm]{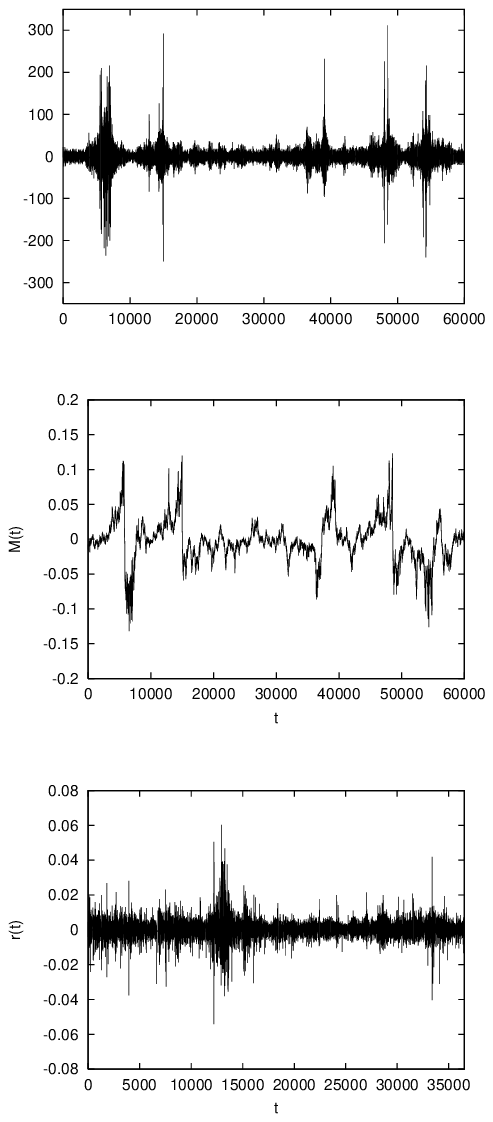}
\caption{(a) Logarithmic return $r(t)=\ln p(t) - \ln p(t-1)$, 
defined as change in magnetization $M(t)$ in the spin model. 
(b) Magnetization $M(t)$ of the spin model.  
(c) For comparison with (a), the log-return of the Dow Jones 
daily changes 1896-1996 is shown.} 
\label{fig1}
\end{figure}

\begin{figure}
\includegraphics[width=14cm]{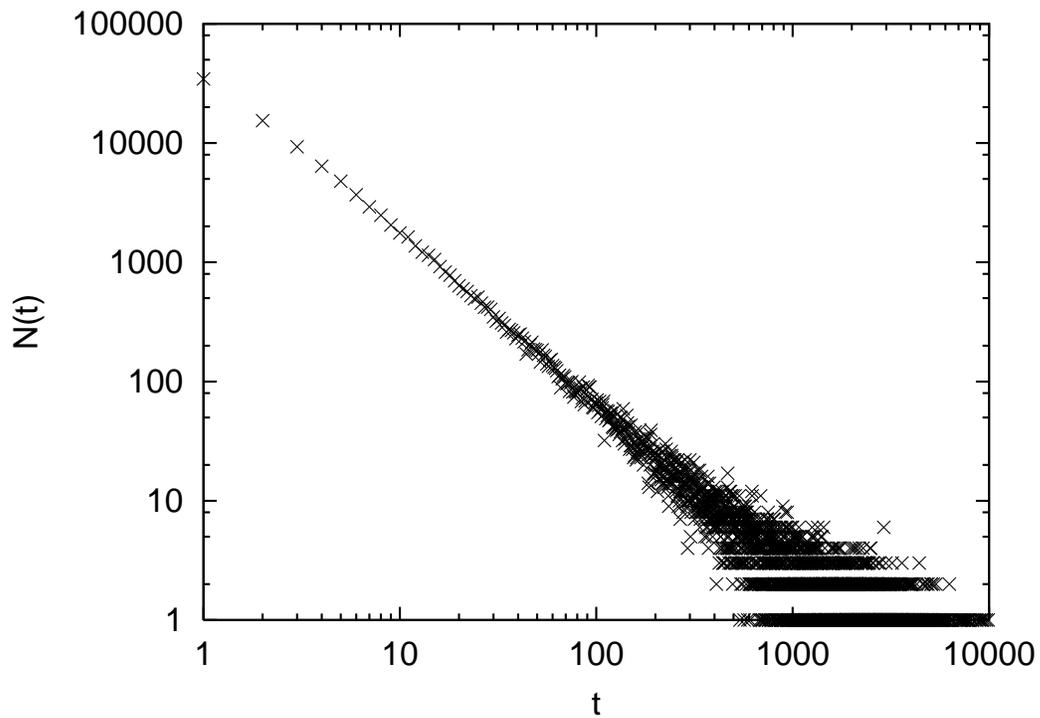}
\caption{Distribution of bull (bear) market durations,
defined as phases with $M(t)>0$ ($M(t) < 0$).} 
\label{fig2}
\end{figure}

\begin{figure}
\includegraphics[width=14cm]{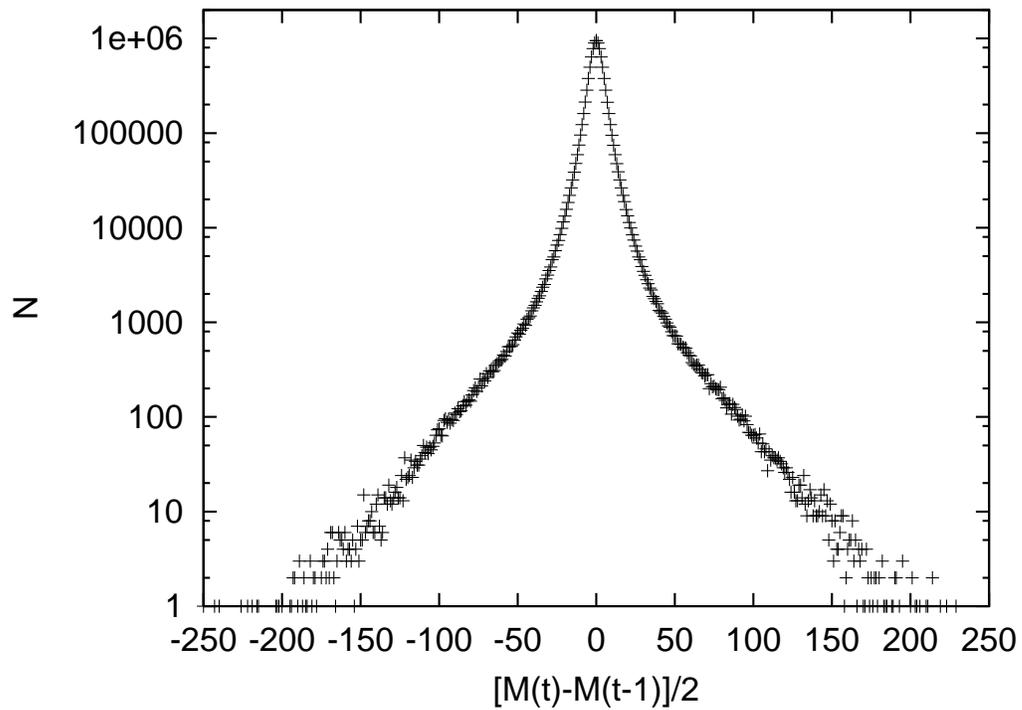}
\caption{Fat tails of the distribution of log-returns.} 
\label{fig3}
\end{figure}

\begin{figure}
\includegraphics[width=14cm]{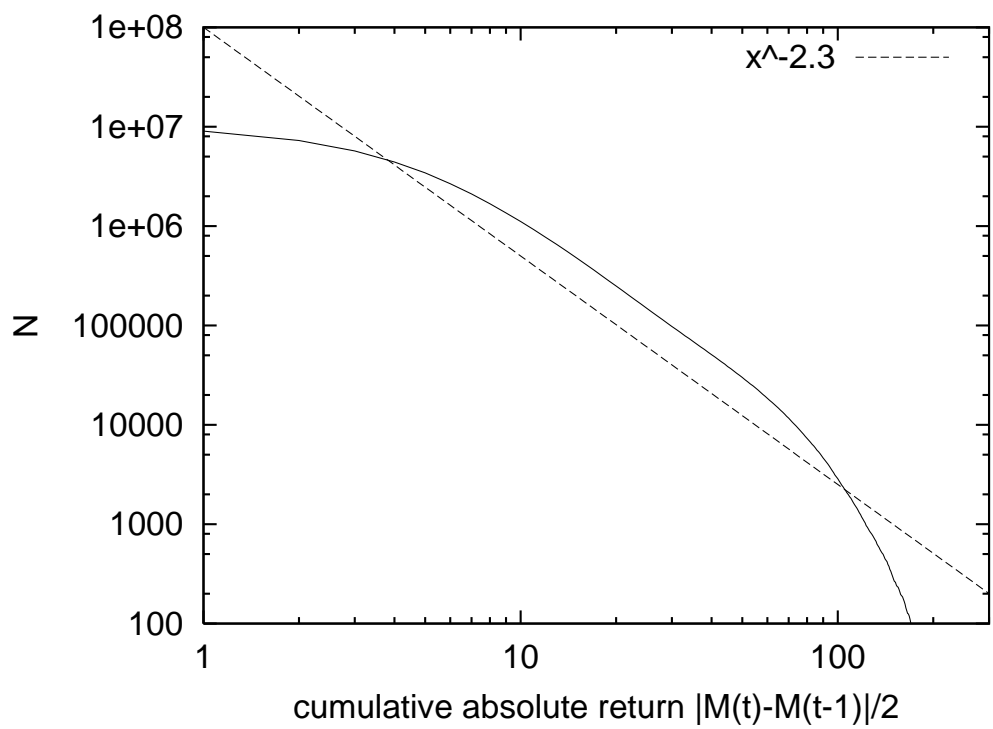}
\caption{Cumulative distribution of log-returns.} 
\label{fig4}
\end{figure}

\begin{figure}
\includegraphics[width=14cm]{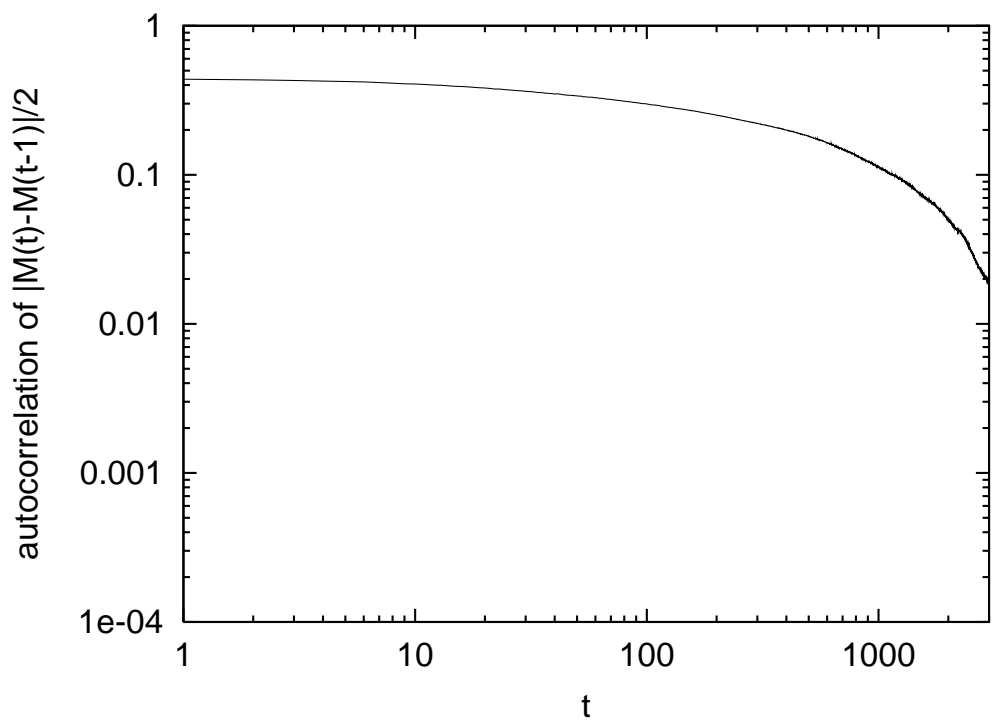}
\caption{Autocorrelation of absolute log-returns.} 
\label{fig5}
\end{figure}

\begin{figure}
\includegraphics[width=14cm]{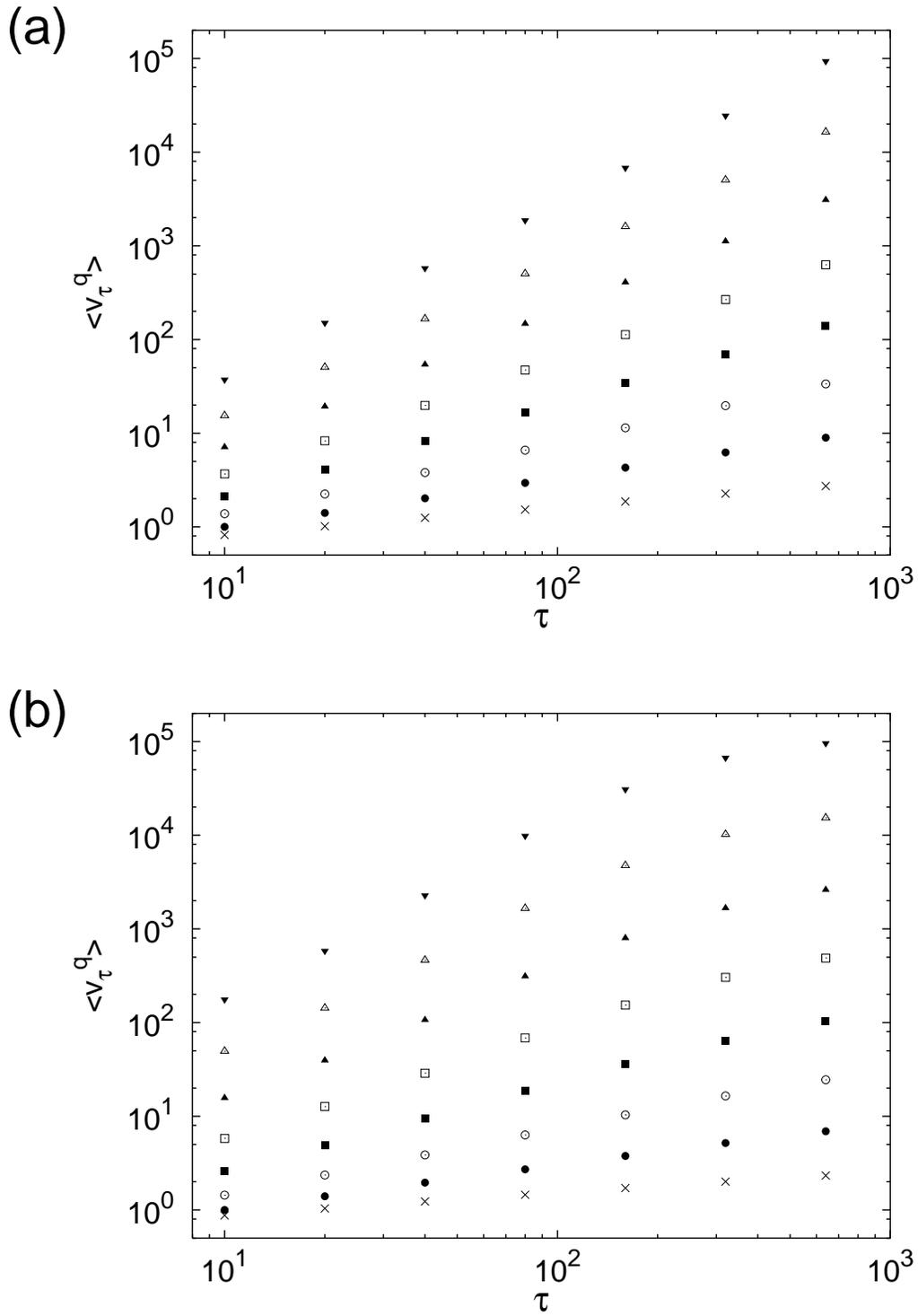}
\caption{%
  Moments ($q$-th order) of volatilities $v_{\tau}$ at time-scales
  $\tau$. (a) For a stock in NYSE with $\tau$ in minutes.
  (b) For the simulation result with $\tau$ in time-steps. In both
  plots, $q$ ranges from 0.5 (cross) to 4.0 (triangles downward) with
  increment 0.5.}
\label{fig6}
\end{figure}

\end{document}